\documentclass[journal]{IEEEtran}

\usepackage{amsmath,amsfonts}
\usepackage{algorithmic}
\usepackage{algorithm}
\usepackage{array}
\usepackage[caption=false,font=footnotesize,labelfont=sf,textfont=sf]{subfig}
\usepackage{textcomp}
\usepackage{stfloats}
\usepackage{url}
\usepackage{verbatim}
\usepackage{graphicx}
\usepackage{subfig}
\usepackage{cite}
\usepackage{booktabs}    % for \toprule, \midrule, \bottomrule
\usepackage{xcolor}      % for color text
\usepackage{array}  
\usepackage{adjustbox}
\usepackage{multirow}
\usepackage{authblk}
\usepackage{soul}
\usepackage{hyperref}    % for clickable links (optional)
\begin{document}

\title{Blockchain for Federated Learning in the Internet of Things: Trustworthy Adaptation, Standards, and the Road Ahead}

\author{
Farhana Javed$^{*}$\thanks{This work was partially funded by Spanish MINECO grants TSI-063000-2021-54 and TSI-063000-2021-55 (6G-DAWN), Grant PID2021-126431OB-I00 funded by MCIN/AEI/10.13039/501100011033, by “ERDF A way of making Europe” (ANEMONE), Generalitat de Catalunya grant 2021 SGR 00770, and by UNITY-6G project, funded from the European Union’s Horizon Europe Smart Networks and Services Joint Undertaking (SNS JU) research and innovation programme under Grant Agreement No. 101192650.},
Engin Zeydan$^{*}$,
Josep Mangues-Bafalluy$^{*}$,
Kapal Dev$^{\dag}$ and
Luis Blanco$^{+}$ \\
$^{*}$Services as Networks (SaS), CTTC/CERCA, Castelldefels, Spain \\
$^{\dag}$Department of Computer Science, Munster Technological University, Cork, Ireland \\
$^{+}$Space and Resilient Communications and Systems (SRCOM), CTTC/CERCA, Castelldefels, Spain \\
\{farhana.javed, josep.mangues, ezeydan, lblanco\}@cttc.es; kapal.dev@mtu.ie
}

% <-this % stops a space


\maketitle

\begin{abstract}
As edge computing gains prominence in Internet of Things (IoTs), smart cities, and autonomous systems, the demand for real-time machine intelligence with low latency and model reliability continues to grow. Federated Learning (FL) addresses these needs by enabling distributed model training without centralizing user data, yet it remains reliant on centralized servers and lacks built-in mechanisms for transparency and trust. Blockchain and Distributed Ledger Technologies (DLTs) can fill this gap by introducing immutability, decentralized coordination, and verifiability into FL workflows. This article presents current standardization efforts from 3GPP, ETSI, ITU-T, IEEE, and O-RAN that steer the integration of FL and blockchain in IoT ecosystems. We then propose a blockchain-based FL framework that replaces the centralized aggregator, incorporates reputation monitoring of IoT devices, and minimizes overhead via selective on-chain storage of model updates. We validate our approach with IOTA Tangle, demonstrating stable throughput and block confirmations, even under increasing FL workloads. Finally, we discuss architectural considerations and future directions for embedding trustworthy and resource-efficient FL in emerging 6G networks and vertical IoT applications. Our results underscore the potential of DLT-enhanced FL to meet stringent trust and energy requirements of next-generation IoT deployments.
\end{abstract}

\begin{IEEEkeywords}
Federated learning, blockchain, Internet of Things (IoTs), standardization, trust, 6G networks.
\end{IEEEkeywords}

%\tableofcontents

\section{Introduction}

\IEEEPARstart{T}{he} Third Generation Partnership Project (3GPP) has been at the forefront of the development of fifth generation (5G) and beyond 5G systems for various applications. In Release 16, 3GPP introduced the Network Data Analytics Function (NWDAF) in its Technical Specifications (TS 23.501 and TS 23.502) to enable data‐driven insights and intelligent decisions in the core network. In Release 17, the analytical scope has been extended to meet the heterogeneous requirements of 5G services. Building on these improvements, Release 18 — often referred to as 5G Advanced — integrates Artificial Intelligence (AI) and Machine Learning (ML) across network operations to enable increasingly flexible and distributed provision of services \cite{Lin2023}. These standardization are in line with the ongoing paradigm shift towards edge computing in the Internet of Things (IoTs) ecosystem. Critical applications in industrial IoT, smart cities, autonomous vehicles, and healthcare require real‐time responsiveness, low latency, and high model reliability. Decentralization of data analytics is critical to meeting these requirements, as edge and fog nodes enable faster, context-aware processing while mitigating the bandwidth constraints of centralized cloud resources. 

Federated Learning (FL) has emerged as a framework for the realization of privacy‐preserving machine intelligence: Instead of collecting raw data in a central repository, FL allows devices to train local models on site and only share parameter updates. This design is particularly attractive for IoT environments characterized by heterogeneous device capabilities, varying data distributions, and intermittent connectivity.

Despite its advantages, FL still faces notable challenges in IoT \cite{FL_Mobile_Network_Management}. Sending large model updates over constrained or intermittent links can increase overhead, and even sharing gradients or parameters can expose sensitive information. In addition, a single, centralized FL aggregator can become a bottleneck and a potential point of failure. For example, the European Telecommunications Standards Institute (ETSI) Zero-touch network and Service Management (ZSM) group has highlighted the need to support decentralized ML services in its report ETSI GS ZSM 012 V1.1.1 (2022-12), underscoring \textit{trustworthiness} and data governance across all management domains. In addition, ETSI TR 104 031 V1.1.1 (2024-01) discusses collaborative AI scenarios, including FL, to work collaboratively without centralizing the data. It emphasizes that decentralized AI/ML services require trustworthy mechanisms to ensure the integrity of the AI/ML models.

The integration of blockchain with FL and IoT can enhance trust and eliminate centralized aggregation servers. In IoT-driven FL, blockchain or Distributed Ledger Technology (DLT) offers a tamper-evident, transparent infrastructure that reduces single points of failure. Contributions such as model parameters and device identifiers are securely recorded and verified, ensuring integrity and minimizing data leakage. 3GPP Release 18 (5G Advanced) extends the 5G System (5GS) with features to monitor FL processes and evaluate participant performance. Meanwhile, ETSI explores DLTs to improve trust in FL ecosystems. Permissioned Distributed Ledger (PDL) technology supports secure, verifiable model exchanges, enabling reliable monitoring of FL participants.

While blockchain can replace the central aggregator in FL for IoT, it also brings challenges. For example, from an IoT perspective, devices cannot absorb additional computing or energy requirements, so blockchain integration must be optimized to avoid resource consumption. In addition, not all data generated by IoT devices is suitable or necessary to send to the blockchain; therefore, it is necessary to determine which data should be uploaded. Given the limitations of IoT devices, gateways are needed to facilitate data transmission to the blockchain. The blockchain itself also has limited resources and must select and manage the data it processes to fulfill the role of aggregator. As the responsibility for aggregation shifts to the blockchain, on-chain verification mechanisms must be established to evaluate the contributions of the models and assign reputation scores to ensure the integrity and quality of the decentralized learning processes.

Several works have explored blockchain and DLTs for decentralized FL. For example, \cite{schmid2020tangle} propose DLT‐based approaches to eliminate the need for a centralized server, while \cite{cao2021dagfl} describes an asynchronous FL framework that uses blockchain for flexible aggregation. In digital twin edge networks, \cite{jiang2022digitaltwin} uses a permissioned blockchain to secure FL updates through smart contracts. Reputation mechanisms are also discussed or implemented in \cite{an2024freb} and \cite{sun2022trust}, which use voting‐based methods to evaluate participant contributions. Further optimizations are proposed in \cite{schmid2020tangle, ren2024flcoin, mazzocca2024feta}, with a focus on scalability and IoT constraints.

Although these works address partial aspects of the overall problem — some remove the central server, others use reputation mechanisms, and still others optimize FL for the IoT — no single approach covers all these aspects at once. Many omit the DLT‐based reputation, retain elements of the centralized design, or overlook the strict resource constraints. A framework that eliminates the central server, takes into account the limitations of the IoT, uses a reputation system, and minimizes overhead remains an open research task. In contrast to previous studies, this paper shows how the integration of blockchain with FL and IoT can enhance trust. The proposed system considers the limitations of blockchain and IoT to optimize resource utilization. It also introduces trust mechanisms that monitor the contributions of IoT devices to protect the quality of the global model. The key contributions are:

\begin{itemize}
    \item From a standardization perspective, we describe the current efforts of Standards Developing Organizations (SDOs) in terms of their contributions and orchestration of the emerging standards and regulatory frameworks that are shaping this convergence.
    \item We propose a framework to enable blockchain-based FL in IoT environments, highlighting a framework where blockchain serves as a replacement for aggregation servers, considering the challenges and limitations of IoTs and blockchain.
    \item We conclude by presenting key findings and a roadmap for the future based on the sixth-generation (6G) network architecture. The integration of trusted AI/ML based on AI/ML and the management of vertical IoT applications based on the proposed work are also provided for a functional architecture.
\end{itemize}

%Addressing performance overhead, Ren et al. \cite{ren2024flcoin} developed FLCoin, a two-layer blockchain system with a committee-based consensus, cutting latency below three seconds and reducing communication overhead by 90\%. Mazzocca et al. \cite{mazzocca2024feta} demonstrated IOTA Tangle’s real-world feasibility for FL through their FETA framework, achieving low power consumption and reduced latency. Schmid et al. \cite{schmid2020tangle} showed that IOTA’s DAG parallel processing improved convergence time, while Lee & Kim \cite{lee2022hybrid} explored hierarchical blockchain-DAG hybrid structures to secure FL with non-IID data.

\begin{table*}[!htbp]
\caption{Summary of Current Standardization Efforts for Blockchain}
\label{tab:standardization}
\renewcommand{\arraystretch}{1.4} % Adjust row spacing
\begin{tabular}{p{3cm} p{2cm} p{6cm} p{4cm}}
    \toprule
    \textbf{Standardization Body} & \textbf{Working Group} & \textbf{Focus Area} & \textbf{Reference / Standard} \\ 
    \midrule
    \multirow{8}{*}{\textbf{ETSI}}   
        & ISG PDL\textsuperscript{*}  & PoC Framework & GS PDL 005  \\
        & ISG PDL\textsuperscript{*}  & Smart Contracts & GS PDL 011  \\ 
        & ISG PDL\textsuperscript{+}  & Landscape of Standards and Technologies & GS PDL 001 \\
        & ISG PDL\textsuperscript{+}  & Application Scenarios & GS PDL 003 \\
        & ISG PDL\textsuperscript{+}  & Inter-Ledger Interoperability & GS PDL 006 \\
    \midrule
    \multirow{12}{*}{\textbf{ITU-T}}  
        & FG DLT\textsuperscript{+} & DLT Terms and Definitions, DLT Overview, Concepts, Ecosystem, Standardization Landscape and Reference Architecture & FG DLT D1.1, D1.2, D1.3 \\
        & FG DLT\textsuperscript{+} & DLT Use Cases & FG DLT D2.1 \\
        & FG DLT\textsuperscript{+} & DLT Regulatory Framework & FG DLT D4.1 \\
        & FG DLT\textsuperscript{¶} & Blockchain \& DLT & REC-F.751.2  \\
        & SG20\textsuperscript{+}   & IoT & Y.dec-IoT-arch, REC-Y.4476  \\
        & SG20\textsuperscript{¶}   & NGN & REC-Y.2342  \\ 
    \midrule
    \multirow{4}{*}{\textbf{IEEE}}   
        & P2144\textsuperscript{+}  & Data Management & Data Management Standards  \\
        & P2418\textsuperscript{+}  & IoT, CAV, Energy & IoT and Energy Standards  \\
        & P2958\textsuperscript{+}  & Identity \& Access Management & Identity and Access Standards  \\
        & P3201-P3214\textsuperscript{+} & Access Control, Interoperability & Access Control Standards   \\
    \midrule
    \multirow{4}{*}{\textbf{CCSA}}   
        & TC1WG6\textsuperscript{*}  & Testing Methods & 2017-0942T-YD  \\
        & TC1WG6\textsuperscript{†}  & Platform (BaaS) & 2019-12527-YD  \\
        & TC5WG6\textsuperscript{†}  & Wireless Network Applications & 2020B896  \\
        & TC10WG1\textsuperscript{‡} & 5G \& Blockchain for IoT & 2020B858  \\ 
    \bottomrule
\end{tabular}

\vspace{5pt}
\noindent
\makebox[\linewidth][l]{%
  \begin{minipage}{1\linewidth}
    \raggedright
    \footnotesize
    \textit{\textbf{Legend:} 
      * PoC \& Testing (PoCs, validation, testing), 
      + Technical Standard (Protocols, security, interoperability), 
      ¶ Reference Framework (Guidelines, architectural frameworks), 
      † Industry Solutions (Application-driven solutions), 
      ‡ Emerging Technologies (Study items for future).
    }
  \end{minipage}%
}
\end{table*}

\section{Overview of Standardization Efforts and Current Challenges}

\subsection{Standardization Efforts for FL in IoT Context}

The 3GPP has prioritized FL in Releases 17 and 18 to enable distributed AI/ML in 5G/6G Radio Access Networks (RAN) and IoT ecosystems. Study Item TR 37.817 of Release 17 defines FL architectures for RAN optimization and introduces components such as the Model Owner, Aggregation Server, and Edge Nodes (User Equipment (UEs), next-generation Node Bs (gNBs), and Distributed Units (DUs)). Release 18 extends FL to network slicing and improves the NWDAF to support FL-driven resource allocation for IoT slices. The Service and System Aspects Working Group 6 (SA6) in TR 33.846 highlights adversarial threats specific to FL in IoT. The main goal is to enable FL-driven RAN optimization while ensuring General Data Protection Regulation (GDPR)-compliant data processing for industrial IoT deployments.

The Open Radio Access Network (O-RAN) Alliance complements 3GPP standards by integrating FL into open RAN architectures. O-RAN Working Group 1 (WG1, Architecture) and Working Group 2 (WG2, AI/ML) focus on FL interoperability within these ecosystems. The O-RAN.WG1.AIML-v02.00 specification embeds FL into the Near-Real-Time RAN Intelligent Controller (Near-RT RIC) and supports FL-based handover tuning for IoT devices via the A1 interface. The O-RAN Service Management and Orchestration (SMO) framework aligns with 3GPP’s NWDAF to enable cross-domain FL workflows for IoT use cases such as smart grid load forecasting. Key challenges include adapting FL to O-RAN’s service-based architecture (SBA) and ensuring low-latency model updates across Distributed Units (DUs). To mitigate security risks like rogue nodes injecting biased gradients, federated attestation mechanisms using hardware root-of-trust modules are proposed.

ETSI’s Experiential Networked Intelligence (ENI) group focuses on FL scalability and regulatory compliance in IoT. ETSI GS ENI 005 highlights challenges in dynamic edge environments, such as fluctuating connectivity, and proposes FL frameworks with lightweight communication to reduce overhead for constrained devices. ETSI TS 103 457 targets cross-border FL deployments, enabling zero-touch workflows for applications like predictive maintenance without centralized oversight. The Institute of Electrical and Electronics Engineers (IEEE) contributes through standards such as IEEE P3652.1 for heterogeneous IoT FL architectures, IEEE P1934.1 for industrial IoT, and IEEE 1451-99 for interoperability via standardized application programming interfaces (APIs). Additionally, the Internet Engineering Task Force (IETF)’s DTLS In Constrained Environments (DICE) working group defines lightweight encryption for FL in IoT, while the Industrial Internet Consortium (IIC) provides reference architectures for real-time asset tracking in logistics.

%These initiatives, alongside 3GPP, O-RAN, ETSI, and IEEE, establish a cohesive foundation for FL in IoT, balancing performance, security, and regulatory compliance. The convergence of these standards enables scalable, privacy-preserving intelligence across next-generation IoT ecosystems, from smart cities to Industry 4.0.

\subsection{Blockchain Standardization Efforts}

A \textit{blockchain} is a well-known form of distributed ledger that organizes data in a sequence of cryptographically linked blocks. Each \textit{block} contains a reference to the hash of its predecessor, which ensures the integrity of the chain. Any attempt to alter a block would modify its hash, making such tampering easily detectable. Several standardization bodies, including the International Telecommunication Union - Telecommunication Standardization Sector (ITU-T), IEEE, ETSI, and China Communications Standards Association (CCSA), are developing specifications and guidelines to ensure interoperability, security, and compliance. Table \ref{tab:standardization} summarizes their main efforts on DLT, including blockchain.

The Industry Specification Group on Permissioned Distributed Ledgers (ISG PDL) of ETSI develops standards and frameworks for permissioned DLT. Documents such as GS PDL 001 outline the technology landscape and standards, while GS PDL 003 explores use cases in the supply chain, finance, and public administration. GS PDL 005 provides guidelines for Proof of Concept (PoC), and GS PDL 011 focuses on smart contract deployment and governance. Additional outputs (GS PDL 002, 004, 007, 012) address architectures, performance metrics, interoperability, and security. The PDL-028 document investigates PDL integration with the oneM2M IoT service layer, focusing on use cases and seamless interworking \cite{ref1}. ETSI GR PDL 032 addresses AI applications in PDL systems and highlights FL as suitable for privacy-preserving decentralized model training in the areas of IoT security, fraud detection, and adaptive learning \cite{ETSI-PDL032}.

ITU-T organizes its blockchain and DLT efforts through Focus Groups (FGs) and Study Groups (SGs). FG DLT produced a series of reports covering terminology, architectures, and standards, while D2.1 outlines use cases in banking, healthcare, and supply chain. Report D4.1 provides regulatory guidance. SG20 focuses on IoT and smart cities, with recommendations such as REC-Y.4476 (IoT data management via blockchain), REC-Y.2342 (next-generation networking), and REC-F.751.2 (secure, standardized blockchain processes).

IEEE defines blockchain standards for diverse applications and interoperability. The P2144 series covers data management and secure exchange; P2418 targets IoT, Connected and Automated Vehicles (CAV), and energy grids; P2958 addresses identity and access management; and the P3201–P3214 suite supports interoperability and access control for healthcare, smart grids, and transportation. In telecommunications, CCSA technical committees develop blockchain standards focused on performance, reliability, Blockchain-as-a-Service (BaaS) scalability, and integration with 5G and IoT, reflecting China’s strategic direction in advanced networks. Additionally, the CAMARA project provides standardized Application Programming Interfaces (APIs) for network services and explores blockchain-like solutions for identity and data provenance in multi-operator environments \cite{Camara-Blockchain}. TM Forum develops DLT-enabled frameworks for billing, roaming, and identity to support blockchain integration with traditional telecom systems \cite{TMForum-TR279}.

\subsection{Challenges}
Consensus mechanisms underlying public blockchains, permissioned frameworks, and Layer-2 (L2) scaling solutions introduce challenges for implementing FL in IoT networks. These mechanisms impact distributed model training across resource-constrained devices, highlighting the need for blockchain designs tailored to IoT environments. While Proof of Stake (PoS) reduces energy consumption compared to Proof of Work (PoW), it still demands significant computational resources and incurs transaction fees, especially under network congestion. Frequent updates to the FL model generate numerous transactions that require immediate validation, which can congest PoS networks and increase latency and costs. In addition, full replication of the ledger on public blockchains exceeds the storage capacity of typical IoT devices.

EVM-based blockchains and smart contracts enable efficient management, but lead to computational overhead and complex maintenance, which limits their practicality in IoT-based FL scenarios. Similarly, permissioned blockchains such as Hyperledger Fabric offer higher transaction throughput but rely on consensus algorithms such as Practical Byzantine Fault Tolerance (PBFT), which increase latency and bandwidth requirements at scale. The iterative nature of FL further exacerbates these scalability limitations. Lightweight blockchain designs must also consider privacy, selective on-chain storage, and continuous verification of the authenticity of devices and model contributions without a centralized aggregator. Achieving FL in IoT therefore requires blockchain designs that balance computational efficiency, storage and verification. %Optimized Byzantine Fault Tolerance, partial on-chain storage and sharding can mitigate resource constraints while ensuring immutability, auditability and verifiability, which are essential for trustworthy FL operations.

\section{Blockchain-based FL in IoT: System Architecture and Workflow}
\subsection{Integration of Blockchain in FL for IoT}

%\fj{Link for the file (page 2): \href{https://drive.google.com/file/d/1YOimnkcmh2UofV8ikY-aAbXiICdqxelJ/view?usp=sharing}{https://drive.google.com/file/d/1YOimnkcmh2UofV8ikY-aAbXiICdqxelJ/view?usp=sharing}.}

%\subsubsection{System Design}

This framework enables end-to-end FL across distributed IoT devices using a DLT-based aggregator for tamper-proof model updates. It ensures privacy, coordinates heterogeneous devices, and supports secure aggregation for vertical applications like business intelligence, reputation scoring, and AI analytics. Key components include:

\textit{IoT Orchestration and Context Manager:} This component manages the interactions between IoT devices, FL processes and blockchain-based validation. It ensures real-time data processing, device trust assessment and adaptive FL workflow optimization. It consists of three modules: IoT Data Analytics \& Insights (IDAI), Trust \& Reputation Management (TRM), and Federated AI \& Optimization Engine (FAOE). IDAI collects and analyzes real-time device data—location, battery, speed—to assess availability for FL participation. TRM scores device trustworthiness based on historical contributions and blockchain-verified compliance, mitigating adversarial risks. FAOE coordinates FL training, device selection, and model updates while optimizing operations based on network conditions and resource constraints. For example, aligned with standardization (e.g., 3GPP’s NWDAF, Policy Control Function (PCF).% this proposed system can leverages real-time insights for resource optimization and enforces decentralized trust policies. Integrating blockchain enables decentralized governance, supporting 3GPP’s security and AI-driven automation for next-generation networks.

\textit{Local Models and IoT Deployed Network:} An IoT network comprises heterogeneous devices collecting domain-specific environmental and operational data for applications like smart cities, healthcare, industrial automation, and traffic management. For example, in traffic scenarios, connected vehicles, cameras, and sensors measure flow rates, speeds, and congestion levels. Each device trains a local model on its own non-independent and identically distributed data, reflecting unique conditions and usage patterns. Local training preserves data confidentiality by eliminating the need to transfer raw data. While this process typically occurs at the device level, similar concepts can apply within the network infrastructure— for instance, a Network Function (NF), such as the Access and Mobility Management Function, could also train local models based on aggregated network data. At this stage, no aggregation occurs—devices or NFs refine models independently based on local distributions. This approach aligns with 3GPP efforts in network data analytics and edge computing, enabling AI-driven optimizations and improved resource allocation through integration with components like NWDAF.%, making the framework flexible for deployment in both 3GPP and non-3GPP IoT environments.

\textit{FL-IoT Decentralized Application (DApp):} This layer operates below IoT orchestration and local model training, enabling secure data exchange and FL aggregation on a distributed ledger. Its core components—the DLT-Adapter, DLT-Verifier, and DLT-Aggregator—are managed on-chain by the DLT-DApp Manager. The DLT-Adapter acts as a gateway, filtering and forwarding essential data (e.g., hashed model weights) from authenticated devices to the blockchain, minimizing on-chain load. The DLT-Verifier scores devices and model updates based on reliability, guiding weighted contributions to improve global model quality. The DLT-Aggregator merges validated updates into the global model, stored on-chain for reference by devices or services. A permissioned IOTA Tangle limits participation reducing the likelihood of malicious actors. These trusted validators confirm transactions faster and help maintain Byzantine fault tolerance, meaning the system can still reach correct consensus even if some validators act arbitrarily or maliciously. Optional milestones—special checkpoints issued by a coordinator—lock in certain points of the Tangle, ensuring transactions are finalized quickly and making the overall process more reliable for sensitive or time-critical applications.

While the previous section outlined the end-to-end architecture for DLT-enabled FL in the IoT, some critical design aspects still need to be considered. In particular, the mechanisms by which IoT devices communicate local model updates via the Tangle, frameworks for reputation and trust management, energy constraints in IoT environments, and the underlying block and transaction data structures that support these operations need to be defined.

\subsection{FL Model Integration of IoT Devices with Blockchain}

Our design leverages a permissioned variant of the IOTA Tangle, inherently reducing consensus overhead. However, direct interaction between IoT nodes and the ledger remains impractical, as repeated confirmations or computations strain energy-limited devices and disrupt critical operations.

\begin{figure*}[htp!]
    \centering
    \includegraphics[width=0.5\textwidth]{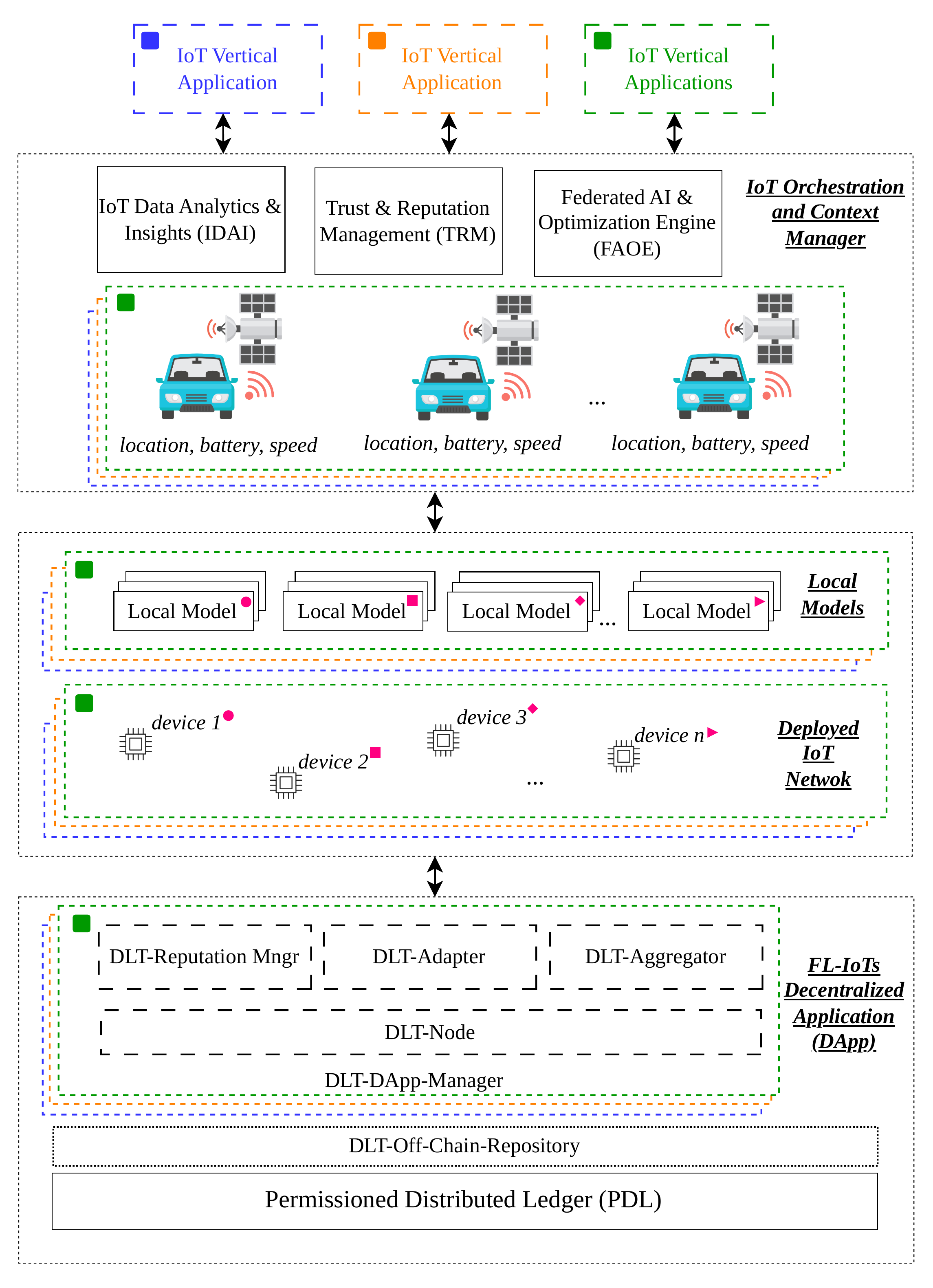}
    \caption{High-level view of system architecture for DLT-enabled trustworthy and FL for IoT.}
    \label{fig:DLT-IoT-FL_Framework}
\end{figure*}

To address this, we embed a DLT-Adapter within the DLT-DApp Manager (Figure \ref{fig:DLT-IoT-FL_Framework}), serving as a verified intermediary between IoT devices and the permissioned Tangle. This approach aligns with ETSI PDL recommendations (ETSI-PDL-028) advocating selective on-chain storage to prevent ledger bloat and reflects 3GPP guidelines on minimizing device-side resource consumption. The DLT-Adapter verifies device identity (informed by the Trust \& Reputation Management module), filters redundant or invalid data, and coordinates further processing of valid local model updates.

In this architecture, IoT nodes transmit local model updates as in standard edge scenarios—without executing any ledger-related computations. The DLT-Adapter receives these updates, shares the model weights securely with the DLT-Aggregator for aggregation, and ensures that only a cryptographic hash of the shared model is anchored on the Tangle. This design preserves model privacy and minimizes on-chain storage, as the actual model weights remain off-chain, accessible only to authorized entities. By recording the hash and a transaction indicating the model's successful sharing, the system ensures immutability and verifiability without incurring unnecessary overhead on IoT devices or exposing sensitive model data on the ledger.

\subsection{Reputation Management}

In this architecture, we also consider that some nodes may be idle, unreliable, or malicious. The \textit{DLT-DApp Manager} consists of two components: the \textit{DLT-Verifier} and the \textit{DLT-Aggregator}. Local model updates are forwarded via the \textit{DLT-Adapter} to the \textit{DLT-Verifier}, which checks the correctness and consistency before recording on- chain. The verifier assigns and updates a reputation score for each node based on the quality of the contributions. Nodes that submit incoherent or outdated updates are penalized, while valid inputs maintain or improve reputation. Reputation scores are recorded on the permissioned ledger to ensure accountability. Nodes whose reputation scores are below a certain threshold are categorized as unreliable, which reduces their influence on future aggregations. The DLT-Verifier updates the reputation of each node by combining its past score with the current model accuracy. A weighting factor controls the influence of past versus current performance. The updated score is recorded on the ledger and used in the aggregation. The \textit{DLT-Aggregator} collects verified updates for the global model computation. Nodes with higher reputation are given more weight, limiting the impact of low-quality or malicious updates. After aggregation, a hash of the global model and references to the contributing updates are committed to the ledger to ensure traceability according to standardized monitoring principles and ETSI PDL trust management guidelines.

\subsection{Lightweight Transactions}

In large-scale, resource-limited IoT environments, it is impractical to store complete model updates in a distributed ledger. Our lightweight transaction framework solves this problem by anchoring only essential metadata and cryptographic hashes on the \textit{IOTA Tangle}, while the actual model data and updates remain off-chain in the repository. The DLT-Adapter acts as an intermediary that authenticates and validates IoT device submissions. Devices can use \textit{MQTT} for low-overhead communication, allowing constrained devices to focus on local model updates without performing ledger-specific operations. Validated updates move to the DLT-Aggregator for global model aggregation, while the adapter anchors only the hashes and minimal metadata on the Tangle using \textit{Zero-Value Transactions}. These carry no fees, and the remote \textit{Proof-of-Work (PoW)} shifts the computational burden from the devices to the permissioned IOTA nodes, saving the overhead of signing transactions for multiple FL rounds. 

%Each FL round involves $n$ devices transmitting updates of size $s$, resulting in $\mathcal{O}(sN)$ communication costs, but selective anchoring keeps on-chain overhead modest.

\section{Experimental Evaluations}

We evaluate the system using two metrics: average transactions per second (TPS) with standard deviation and variability, and block processing time—defined as the delay between transaction submission and confirmation by a milestone in IOTA. Each transaction represents a model update from a simulated IoT client in an FL round. Experiments ran for 10, 30, and 50 rounds, repeated 10 times for consistency, on Ubuntu 22.04 LTS with an i7-1265U CPU (10 cores, up to 4.8 GHz) and 31 GiB RAM. A private, permissioned IOTA Tangle (Hornet v2.0.2) with two nodes handled milestones and validation. Node 1 received transactions via the DLT-Adapter; Node 2 managed gossip and confirmations. An INX coordinator handled milestones, and monitoring was included. All components ran in Docker (details in Table~\ref{tab:simulation_setup}). Transactions were sent using a Python client through a local Mosquitto MQTT broker (v2.0.11) and Paho-MQTT (v2.1.0). Payloads ranged from 1.5–3 KB, aligned with IOTA’s 32 KB limit, and weights were stored off-chain via IPFS (v0.21.0). The FL setup used public IoT data\footnote{\url{https://iotanalytics.unsw.edu.au/iottraces.html}}, partitioned across 20 simulated clients, each training a lightweight neural network (one hidden layer) for 20 local epochs. Updates were aggregated using FedAvg, with reputation scores based on validation accuracy.

\begin{table}[tp!]
\centering
\caption{Simulation environment setup}
\begin{tabular}{p{2.5cm} p{4.5cm}}
\toprule
\textbf{Host OS} & Linux (Ubuntu 22.04 LTS) \\
\midrule
\textbf{CPU} & 12th Gen Intel(R) Core(TM) i7-1265U, 10 cores, max frequency 4.8 GHz \\
\midrule
\textbf{Memory} & 31 GiB DDR4 \\
\midrule
\textbf{Docker Containers} & 
Hornet Node 1: CPU 0.59\%, 583.9 MiB \newline
Hornet Node 2: CPU 3.13\%, 581.6 MiB \newline
INX Coordinator: 10.49 MiB \newline
Traefik (proxy): 21.27 MiB \newline
Monitoring and explorer services deployed \\
\midrule
\textbf{Tangle Setup} & 
Hornet v2.0.2 \newline
Base token: SandCoin (SAND) \newline
Milestone interval: default 10 s \newline
Block rate: $\sim$0.2 blocks/s \\
\midrule
\textbf{Payload Sizes} & 
DLT-Adapter: 2--3 KB \newline
DLT-Aggregator: 2--3 KB \newline
DLT-Verifier: 1.5--2 KB \\
\midrule
\textbf{MQTT Setup} & 
Local broker (Mosquitto v2.0.11) on \texttt{localhost:1883} \newline
Paho-MQTT Python client (v2.1.0) used for publishing \\
\midrule
\textbf{Off-chain Storage} & 
IPFS (v0.21.0) \\
\bottomrule
\end{tabular}
\label{tab:simulation_setup}
\end{table}

\subsection{Simulation Results}

Table~\ref{tab:fl_throughput_variability} shows system throughput and variability metrics for FL executions consisting of 10, 30 and 50 rounds. These results provide important insights into the scaling of transaction volume and node performance as FL update bursts increase, an important consideration for balancing low-latency responses and frequent model updates in IoT deployments. As shown in Table~\ref{tab:fl_throughput_variability}, the average processed transactions per second (TX/sec) increase from 1.82 at 10 rounds to 2.10 and 2.12 at 30 and 50 rounds, respectively. Parallel to this increase in throughput, the standard deviation and variability fall sharply — from 30.8\% at 10 rounds to 7.25\% at 30 rounds and 4.9\% at 50 rounds. This trend reflects an important observation: the system becomes more stable and predictable when it operates under higher, sustained load conditions. This stability stems from the fact that a well-configured IOTA node can accommodate predictable bursts without congestion or erratic queue buildup, confirming the resilience and scalability of the system for moderate- to large IoT FL workloads.

\begin{table}[tp!]
\centering
\caption{System Throughput and Variability During FL Execution}
\begin{tabular}{lccc}
\toprule
\textbf{FL Rounds} & \textbf{Average TX/sec} & \textbf{Std Dev} & \textbf{Variability (\%)} \\
\midrule
10                 & 1.82                   & 0.56             & 30.8\%                    \\
30                 & 2.10                   & 0.15             & 7.1\%                     \\
50 & 2.12 & 0.10 & 4.90\% \\
\bottomrule
\end{tabular}
\label{tab:fl_throughput_variability}
\end{table}

\begin{figure}[tp!]
    \centering
    \includegraphics[width=0.9\columnwidth]{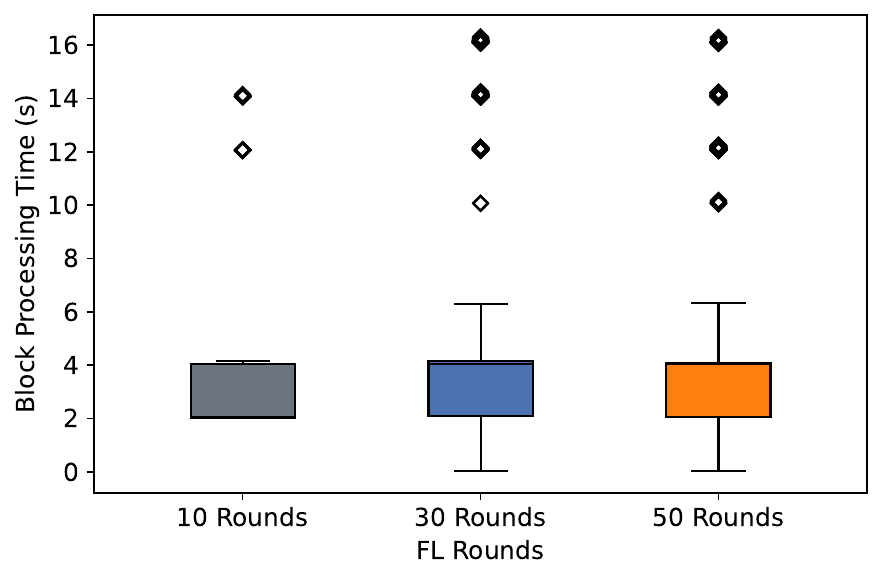}
    \caption{Distribution of block processing time across different FL rounds—10, 30, and 50.}
    \label{fig:block}
\end{figure}

Figure~\ref{fig:block} presents the distribution of block processing times for each FL scenario. Interestingly, we do not observe a significant upward shift in block processing times as the number of FL rounds increases. Instead, the block processing times remain consistent in all three scenarios, with median values clustering around 2 to 4 seconds and occasional outliers in the range of 10 to 16 seconds. These outliers correspond to the  fixed milestone issuance intervals of nodes (approximately 10 seconds) and reflect transactions that just miss one milestone and wait for the next cycle. The fact that block processing times did not increase at higher loads indicates that the system has made good use of node processing capacity with no sustained backlog or congestion. This behavior underscores that block confirmation times are driven more by milestone issuance rates than by incremental increases in transaction volume — as long as the system remains within operational limits. Furthermore, the low variability and consistent processing times observed at higher FL rounds confirm the robustness of the node in handling sustained FL workloads.

The observed behavior provides a realistic baseline for IOTA-based FL deployments in larger setups, factors such as network propagation delays, random transaction spikes, and distributed milestone confirmations can lead to additional fluctuations. Nonetheless, our results show that the setup remained stable and responsive at moderate to high transaction volumes (up to 50 FL rounds with 20 devices each), confirming its adaptation for FL scenarios. \\
%These experiments demonstrate that the private IOTA Tangle can handle FL workloads from 10 to 50 rounds in IoT scenarios. The results indicate that milestone intervals and network size are critical for applications requiring real-time guarantees. Aligning FL updates with milestone scheduling and adjusting network parameters can support reliable ledger operations and distributed learning in IoT deployments.

%Overall, these extensions serve as a natural evolution of our current PoC. While the existing setup demonstrates the viability of integrating FL with IOTA on a modest scale, future work could involve more comprehensive testing with a larger node base and higher transaction volumes, potentially including thousands of IoT clients over geographically dispersed Tangle networks. Taken together, these steps would not only validate our design choices but also carve out a path for fully realizing a secure, low‐latency, and scalable distributed ledger framework for federated learning in resource‐constrained IoT ecosystems.

\begin{figure*}[htp!]
    \centering
    \includegraphics[width=.7\textwidth]{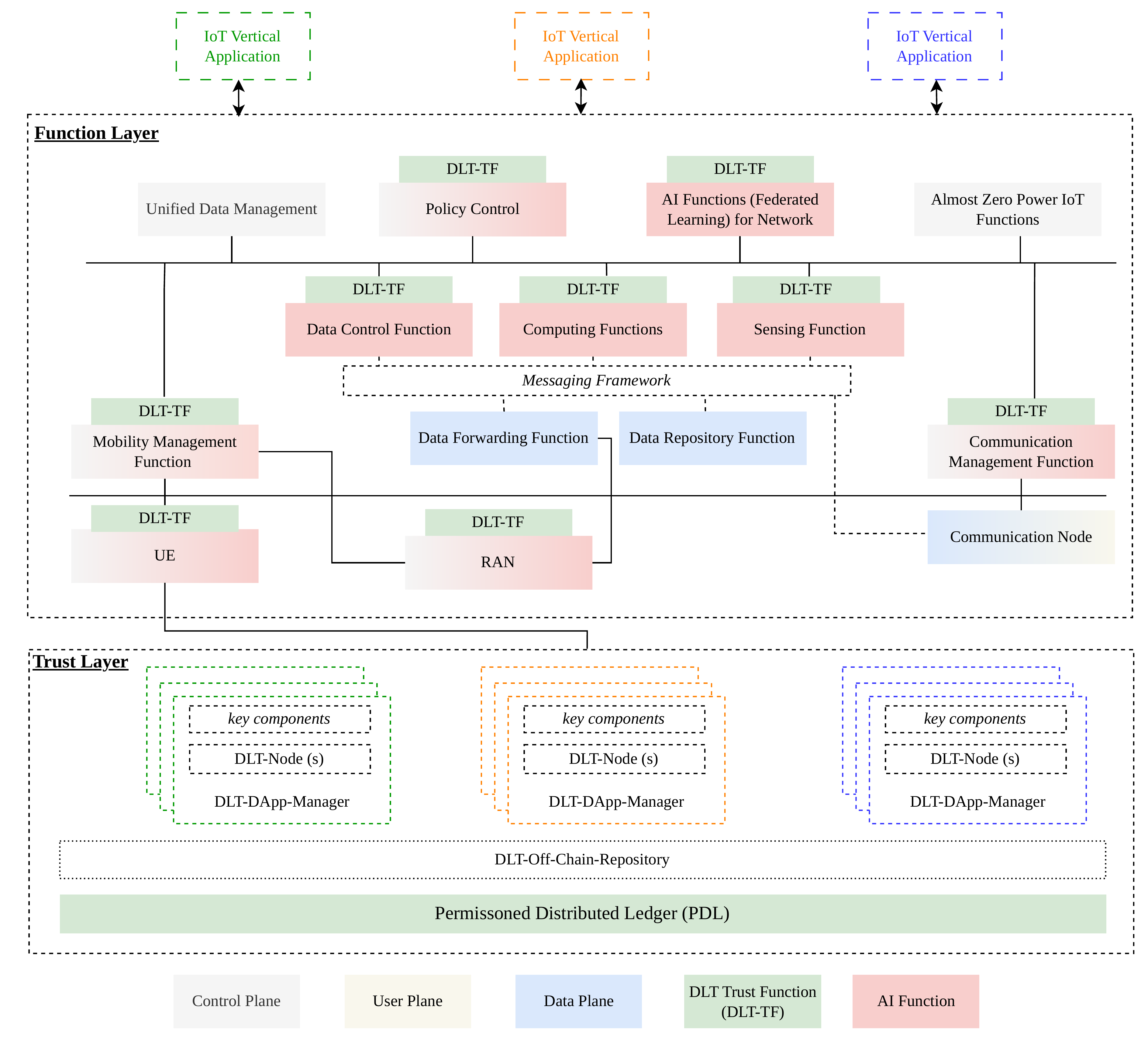}
    \caption{6G network architecture integrating  DLT, AI for trustworthy 6G networks services. The framework illustrates key functional modules, decentralized trust and orchestration components, and support for IoT vertical applications across user, control, and data planes.}
    \label{fig:DLT-FL-IoT-6G}
\end{figure*}

\section{Discussion and Future Directions}
\subsection{Discussion and Key Takeaways}

\textit{Improving Blockchain Network Scalability:} A key takeaway from our results is the importance of scalability optimization when integrating FL with DLTs such as IOTA. Our proof-of-concept processes each FL update as an individual transaction hash on the ledger. While this approach is effective and stable at moderate scale, as confirmed by our PoC, transaction volumes will inevitably increase with larger federations. In such cases, \textit{batch processing} can aggregate multiple FL updates into single transactions, reducing ledger congestion and minimizing queuing delays. However, determining the optimal batch size is essential to avoid latency for individual contributions. Scaling beyond small distributed setups will also introduce additional overhead from gossip propagation and milestone synchronization, suggesting that future architectures should consider hierarchical clustering or multi-layered node designs. Although our current experiments did not encounter severe congestion, it is anticipated that under extreme burst conditions, bottlenecks in the gossip layer and milestone scheduling could arise, underscoring the need for proactive resource planning and dynamic scheduling strategies.

\textit{Latency-Security Trade-offs:} IOTA’s milestone-based confirmation mechanism delivers strong consistency and data integrity for FL model updates but naturally introduces confirmation latency, particularly for transactions that narrowly miss milestone intervals. While this latency remained predictable in our controlled environment, larger and more dynamic deployments may benefit from selective mechanisms such as partial confirmations or fast-lane processing for time-sensitive IoT data. This highlights a trade-off between latency and integrity that becomes increasingly significant for real-time or mission-critical decision-making in large-scale FL deployments.

\textit{Energy Efficiency and Off-Chain Extensions:} In resource-constrained IoT settings, energy efficiency and off-chain solutions are vital. Our setup simulated devices communicating through an MQTT broker, with payloads sized to reflect realistic FL model updates (200 KB to 2 MB). However, transmitting these updates over constrained wireless links (such as NB-IoT) would amplify energy costs and tail latencies. Solutions that rely on off-chain storage such as IPFS can reduce the complexity of on-chain data and the transaction load. As systems scale, future architectures may need to incorporate side-DAGs, sharding techniques, or advanced off-chain protocols to balance computational overhead, data availability, and ledger integrity while remaining energy-aware.

\subsection{Future Directions}

Figure \ref{fig:DLT-FL-IoT-6G} outlines a 6G network architecture that integrates AI, FL and DLT. The Function Layer (e.g. DLT-TF modules for Unified Data Management, AI/FL and Sensing Functions) emphasizes native intelligence and distributed computing and supports IoT services such as smart cities and Industry 4.0. The Trust Layer (DLT nodes and DApp-Manager ) and the PDL Layer (integration of control/user/data planes with DLT and AI) emphasize decentralized security and transparency, which are essential for FL in IoT ecosystems. While this is in line with the 6G vision of trustworthiness \cite{NGMN2023Trust} as well as converged communications, sensing, and computing \cite{vivo6G}, there are still gaps in scalability, interoperability and standardization.

3GPP is extending 5G standards to support native AI, FL frameworks and ISAC capabilities, with the proposed DLT solution serving as a trust layer for blockchain-based FL and AI orchestration. Nevertheless, there are some challenges that hinder adoption. First, scaling DLT-based FL auditing still faces bottlenecks in resource-constrained IoT. Second, fragmented deployments arise without unified standards for DLT-FL interoperability. Third, the lack of FL-specific protocols for AZP IoT devices — including energy-efficient model training— - hinders their widespread use in battery-powered scenarios. These challenges make it clear that standardisation bodies such as 3GPP and ETSI’s PDL working group need to develop adapter-centric frameworks, lightweight identity management and IoT-specific DLT mechanisms.

To solve these problems, standardization must focus on four pillars: First, lightweight consensus mechanisms (e.g. delegated PoS) and hierarchical architectures (e.g. parent-child chains) can optimize FL rounds without compromising auditability. Second, unified interoperability protocols (e.g. standardized APIs) are essential for the integration of DLT-TF with 3GPP/O-RAN functions. Third, energy-aware FL standards must include adaptive participation thresholds and model compression tailored to AZP-IoT. Finally, the definition of DLT performance metrics — such as TPS and latency benchmarks for FL-IoT — can ensure alignment with ITU-T guidelines and promote practical applicability.

%By resolving these priorities, the vision of a secure, intelligent 6G ecosystem—where DLT underpins trust in FL-driven IoT applications—can transition from architectural promise to operational reality.

%Future steps include developing interoperable DLT frameworks under 3GPP and O-RAN standards, optimizing DLT consensus mechanisms for IoT devices, and designing lightweight protocols for real-time AI workflows. Ensuring regulatory compliance, such as GDPR, is crucial. Practical use cases in healthcare, smart cities, and Industry 4.0 will validate the technology, driven by collaboration among telecom operators, AI experts, and blockchain innovators.

%\fj{Link for the file (page 1): \href{https://drive.google.com/file/d/1YOimnkcmh2UofV8ikY-aAbXiICdqxelJ/view?usp=sharing}{https://drive.google.com/file/d/1YOimnkcmh2UofV8ikY-aAbXiICdqxelJ/view?usp=sharing}.}

\section{Conclusion}
This article summarizes the standardization efforts of key Standards Developing Organizations (SDOs)—including 3GPP, ETSI, ITU-T, IEEE, and O-RAN—that drive Federated Learning (FL) and blockchain convergence in the Internet of Things (IoT). These coordinated initiatives underscore how Distributed Ledger Technologies (DLTs) empower secure, decentralized intelligence in next-generation networks.

We also address the challenges inherent in blockchain-based FL for IoT and propose a framework that eliminates centralized aggregators by leveraging DLT-based components. Our system facilitates verifiable contributions, reputation-driven data integrity, and accommodates IoT resource constraints. Finally, we explore future directions for embedding native trust within network architectures. Trust must be integral to FL, via decentralized model verification and alignment with evolving standards and regulations. We envision the synergy of blockchain, FL, and IoT—fortified by interoperability and built-in trust—as pivotal for scalable, secure future 6G and industrial applications.

%\section*{Acknowledgments}

%\section{Simple References}

\newpage

\vfill

\end{document}